\documentclass{article}

\usepackage{PRIMEarxiv}

\usepackage[utf8]{inputenc} % allow utf-8 input
\usepackage[T1]{fontenc}    % proper font encoding
\usepackage{hyperref}       % hyperlinks
\usepackage{url}            % simple URL typesetting
\usepackage{booktabs}       % professional-quality tables
\usepackage{amsfonts}       % blackboard math symbols
\usepackage{nicefrac}       % compact symbols for 1/2, etc.
\usepackage[expansion=false]{microtype} % microtypography
\usepackage{graphicx}       % graphics
\graphicspath{{Images/}}
\usepackage{amsmath}        % math environments and \text
\usepackage{subcaption}
\usepackage{float}
\usepackage[section]{placeins}

\title{Cyber Risk Scoring with QUBO: A Quantum and Hybrid Benchmark Study}

\author{
  Remo Marini \\
  \textit{Assicurazioni Generali S.p.A., Trieste, Italy} \\
  \textit{Rewrite Technology, Milan, Italy} \\
  \textit{F3RM1 Foundation, Milan, Italy}
  \And
  Riccardo Arpe\thanks{Corresponding author: \texttt{riccardo@rewrite.technology}} \\
  \textit{Rewrite Technology, Milan, Italy} \\
  \textit{F3RM1 Foundation, Milan, Italy}
}

\begin{document}
\maketitle

\begin{abstract}
Assessing cyber risk in complex IT infrastructures is challenging due to the dynamic, interconnected nature of digital systems. Traditional methods rely on static, qualitative models that fail to scale and ignore systemic interdependencies. We introduce a quantitative approach to cyber risk assessment based on Quadratic Unconstrained Binary Optimization (QUBO), a formulation compatible with both classical computing and quantum annealing. Applied to a realistic 255-node layered infrastructure, the model reveals how risk spreads along non-trivial propagation paths invisible to visual inspection. We further conduct scalability experiments on networks up to 800 nodes, comparing classical, quantum, and hybrid workflows. Although quantum annealing produces solutions comparable to classical heuristics, its advantages are hindered by the embedding overhead required to map dense cyber-risk QUBOs onto current quantum hardware. Hybrid quantum-classical solvers avoid this bottleneck, combining competitive scaling with superior exploration of the solution space and more stable risk configurations. This work delivers two advances: a rigorous, tunable, and generalizable QUBO model for cyber risk, adaptable to diverse infrastructures through flexible parameterization; and the first comparative study of classical, quantum, and hybrid solvers for cyber risk scoring at scale.
\end{abstract}


\section{Introduction}

Modern societies rely heavily on digital and cyber infrastructures, from enterprise IT networks to industrial control systems and critical services. The growing size, interconnectivity, and heterogeneity of these infrastructures make cyber risk assessment both essential and highly challenging. Moreover, cyber risk is inherently dynamic: as new vulnerabilities and attack strategies emerge, static assessments quickly become outdated \cite{cheimonidis2023}. This tension between protecting vital systems and quantifying evolving threats drives the need for rigorous, scalable, and adaptive risk models.

Traditional risk-scoring methods face significant limitations. Many organizations rely on qualitative checklists or subjective ratings (low/medium/high), which are inconsistent across assessors and prone to bias \cite{simpson2024}. Standardized scoring systems such as CVSS \cite{mell2007} provide per-vulnerability severity ratings but do not account for network topology or inter-component propagation. Even quantitative frameworks, such as those proposed in enterprise risk management or compliance audits, often capture only asset-level snapshots, ignoring the interdependencies between components. Furthermore, external scoring approaches, such as enterprise cyber ratings, have been shown to correlate poorly with actual incident data \cite{riskwriter2018,eling2019}, reinforcing concerns about their reliability. Yet, real-world infrastructures display precisely the opposite behavior: cyber risk emerges from the interaction of components, spreading across control and information flows in ways that are often counterintuitive \cite{kavallieratos2020,kavallieratos2021}. Without capturing these interdependencies, assessments underestimate systemic vulnerabilities and fail to anticipate cascading failures.

Recent research has emphasized the propagation and dependence of cyber risk across networked systems, showing how localized vulnerabilities can compound through interdependencies \cite{da2021}. These insights highlight the limitations of static or purely linear models, which cannot capture second-order interactions, multi-hop propagation, or the influence of connectivity on systemic stability. At the same time, classical optimization and simulation approaches struggle with scaling: as infrastructures grow, the number of interdependencies explodes, making brute-force modeling computationally prohibitive.

To address these challenges, we propose a fully quantitative risk model based on Quadratic Unconstrained Binary Optimization (QUBO). QUBO encodes optimization problems using binary variables with a quadratic cost function, naturally representing both local contributions and pairwise interactions. In our formulation, node-specific risk factors --- including patching status, internet exposure, and initial risk scores --- are combined with connectivity-based penalties and neighbor-influence terms into a Hamiltonian. Minimizing this Hamiltonian yields the most likely systemic risk configuration. Unlike conventional models, this approach explicitly encodes how vulnerabilities propagate and interact, producing a dynamic and emergent view of network-level risk.

A central advantage of QUBO is its flexibility. Practitioners can construct arbitrary starting networks with heterogeneous nodes, adjust Hamiltonian terms to reflect industry-specific risks, and fine-tune the weighting coefficients to emphasize different propagation dynamics. This adaptability makes the approach applicable across a wide range of domains, from small IT networks to enterprise-scale infrastructures, and even to forms of risk beyond cybersecurity. In contrast to current qualitative or snapshot-based methods, our framework offers a rigorous, tunable, and consistent quantitative alternative.

Our goal in the following experiments is twofold: first, to show how the model behaves across infrastructures of increasing size and complexity; second, to understand how different solvers navigate the optimization landscape and whether the resulting solutions are fundamentally stable or prone to escalation.

We demonstrate this framework on small benchmark networks, a realistic 255-nodes layered infrastructure, and extended tests on IT environments of up to 800 nodes. Results show that risk tends to concentrate on central, highly interconnected nodes, spreading along non-trivial paths that are invisible to visual inspection. To better understand the behavior of the optimization landscape, we also apply the QUBO minimization recursively: although not standard practice, this iterative procedure provides valuable insight into the stability of the obtained solutions. In stable cases, repeated minimization produces essentially the same outcome, showing that the network settles into a balanced configuration. In unstable cases, the risk keeps increasing at each iteration, revealing that the solver has reached a shallow or fragile region of the energy landscape instead of a robust minimum.

To address computational challenges and investigate scalability, we compare classical metaheuristics such as Tabu Search \cite{Glover1989,Glover1990} with quantum annealing and hybrid quantum--classical workflows. Quantum annealing has been highlighted as a promising approach for tackling dense and rugged optimization landscapes \cite{Hauke2020,yarkoni2021}, and recent studies have begun exploring its relevance for cybersecurity applications \cite{carney2022}. In the work that follows, we evaluate the strengths and limitations of each solution workflow, quantitatively assessing their solution quality and scalability. The ultimate goal is to identify a practical and effective tool for analyzing real-world IT infrastructures and, eventually, to support large-scale deployment of QUBO-based cyber risk assessment.

\section{Related Work}
\label{sec:RelatedWork}

Quantitative cyber risk assessment has been approached from several directions, each with distinct strengths and limitations. Standardized frameworks such as NIST SP 800-30 \cite{nist80030} provide structured methodologies for risk identification and evaluation but rely on categorical ratings and expert judgment, limiting their ability to capture systemic interdependencies. The Factor Analysis of Information Risk (FAIR) model addresses this in part by decomposing risk into quantifiable factors such as loss event frequency and magnitude, often combined with Monte Carlo simulation to produce probabilistic estimates. Wang et al.\ \cite{wang2020} extended the FAIR model using Bayesian networks to reason about interdependencies between risk factors, demonstrating improved expressiveness over the original taxonomy. Monte Carlo methods have also been applied to estimate aggregate cyber loss distributions \cite{bohme2010}, though these approaches typically model risk at the portfolio or organizational level rather than at the infrastructure topology level. Overall, FAIR-based approaches remain focused on individual assets or scenarios and do not natively model risk propagation across network topologies.

Graph-based methods offer a complementary perspective. Bayesian attack graphs encode possible attack paths and allow probabilistic inference about network compromise \cite{poolsappasit2012}. Zeng et al.\ \cite{zeng2019} provide a comprehensive survey of attack graph analysis methods, including Bayesian, Markov, and cost-optimization approaches, while Zenitani \cite{zenitani2023} offers a recent explanatory guide to attack graph techniques. These models are effective at capturing multi-step attack progression and have been applied to dynamic risk management. However, they require explicit enumeration of attack paths and vulnerability preconditions, which limits scalability as infrastructure complexity grows. Separately, epidemic models inspired by the SIR framework have been adapted to study malware propagation and cyber risk spreading across networks, treating compromise as an infection process. While these approaches capture temporal dynamics, they typically assume homogeneous node populations and do not account for the heterogeneous risk profiles and asymmetric connectivity found in real IT infrastructures.

More recently, quantum computing has been explored as a tool for security-related optimization. Kritsadakul and Kantabutra \cite{kritsadakul2025} formulated layered cybersecurity strategies as QUBO problems suitable for quantum annealing, providing correctness proofs and experimental validation on D-Wave simulators. In a preprint, Carney \cite{carney2022} investigated QUBO formulations for modeling kill-chain disruption using quantum computing. These studies demonstrate the feasibility of encoding security problems into quantum-compatible formats but do not address systemic risk scoring across full IT infrastructures or provide comparative benchmarks across solver types.

Our work differs from existing approaches in two key respects. First, we formulate cyber risk assessment as a single QUBO whose Hamiltonian jointly encodes local vulnerability attributes, connectivity penalties, neighbor influence, and exposure flags, producing an emergent risk configuration rather than enumerating individual attack paths. Second, we provide a direct comparison of classical, quantum, and hybrid solvers on the same risk-scoring problem at scale, benchmarking not only solution quality but also stability under recursive minimization.

\section{Methods}
\label{sec:Methods}

\subsection{Infrastructure Model}

\begin{figure}[!htbp]
    \centering
    \includegraphics[width=0.8\linewidth]{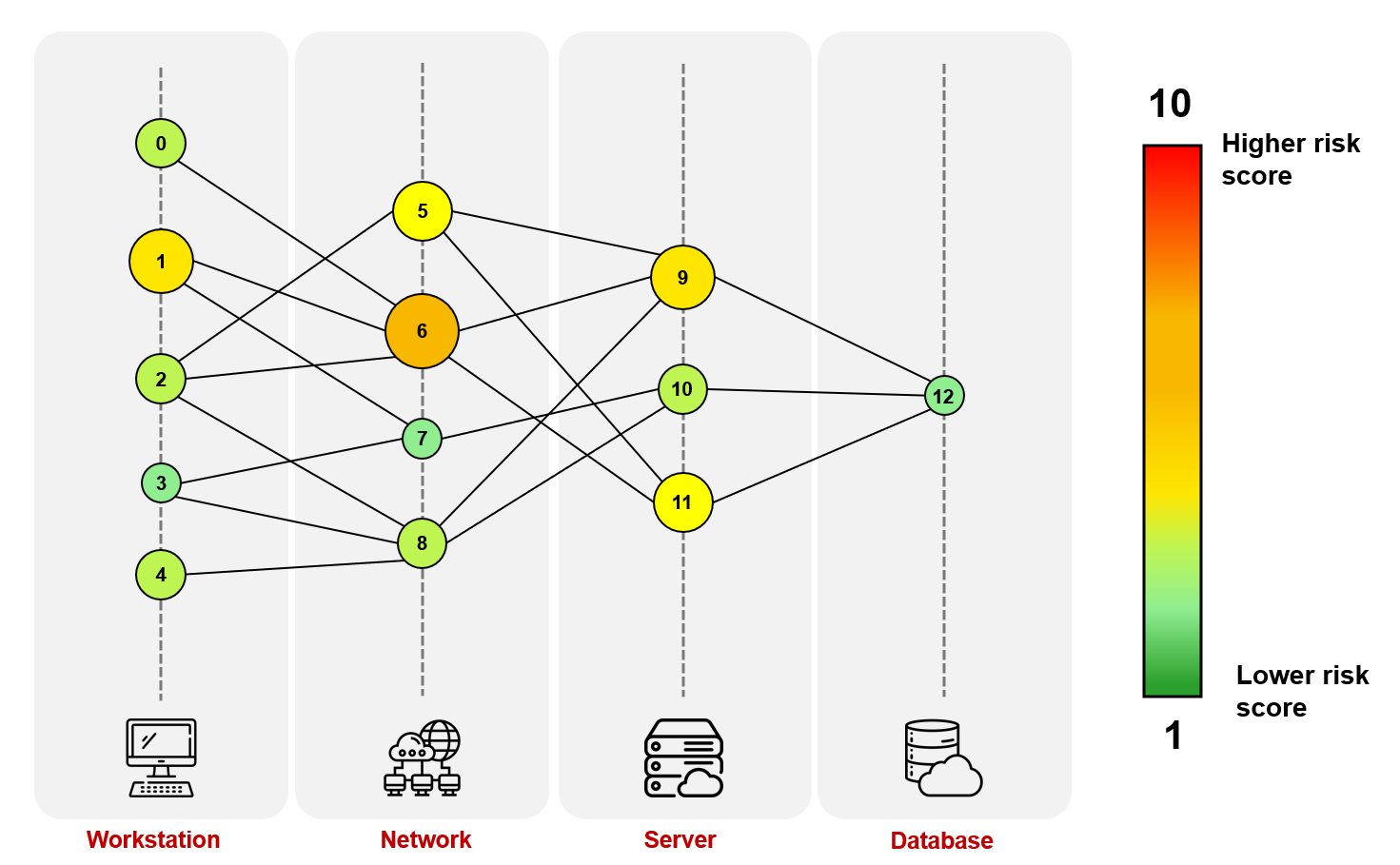}
    \caption{Test network topology. Nodes are grouped into four layers: workstations, network, servers, and databases. Color and size depend on the initial risk score, while numbers reflect the node identification.}
    \label{fig:topology}
\end{figure}

To validate our QUBO-based cyber risk modeling framework, we employed a synthetic yet realistic layered IT architecture, depicted in Fig.~\ref{fig:topology}. The infrastructure is structured across four distinct layers: workstations, network components, servers, and databases. Each node in the graph represents an individual system component, and edges indicate bilateral logical or physical communication links between them.

Every node is assigned an initial risk score ($IS$) ranging from 1 to 10. These scores reflect intrinsic vulnerabilities due to poor maintenance, high exposure to external threats, or critical operational roles. A key aspect of this approach is that the initial score of each node should represent a well-thought-out, isolated risk evaluation--independent of its position or connections within the network. This is precisely where our QUBO model becomes valuable: when risk owners of various assets assign local risk values, our model can extrapolate the overall systemic risk that emerges from interconnectivity and propagation across the network.

The initial risk of each node is visually represented in the topology by both the size of the circle (larger for higher risk) and its color (green to red gradient indicating increasing risk). This dual encoding provides an immediate visual sense of the starting conditions before any propagation effects are computed.

Each node also carries a binary \textit{update status flag} ($f^{no\_update}$ - indicating whether it receives regular security updates) and an \textit{exposure attribute} ($f^{internet}$ - flagging systems exposed to internet). These attributes act as implicit multipliers or thresholds within the Hamiltonian components, as described in Section~\ref{sec:ProblemFormulation}. They modulate both individual risk and propagation behavior in the optimization landscape.

The graph structure is fully flexible: nodes can be added or removed, edge weights can be tuned to reflect different communication intensities, and new layers can be incorporated. Although our method supports general \( n \)-dimensional graph embeddings, we adopted a 2D layered layout here for visual clarity and simplicity of interpretation. This layout also facilitates qualitative validation of risk propagation patterns during early stages of model calibration.

We focused on this simple, interpretable scenario as a benchmark to tune our QUBO model. The model's effectiveness depends not only on the mathematical formulation of the Hamiltonians but also, crucially, on the tuning of their relative weights (the \(\lambda_i\) coefficients). Having a comprehensible and structured network was essential to ensure that the model's outputs aligned with expectations from domain experts. In this way, the benchmark scenario acted as a calibration platform for both the design and fine-tuning of our cyber risk QUBO.

\begin{figure}[!htbp]
    \centering
    \includegraphics[width=0.9\linewidth]{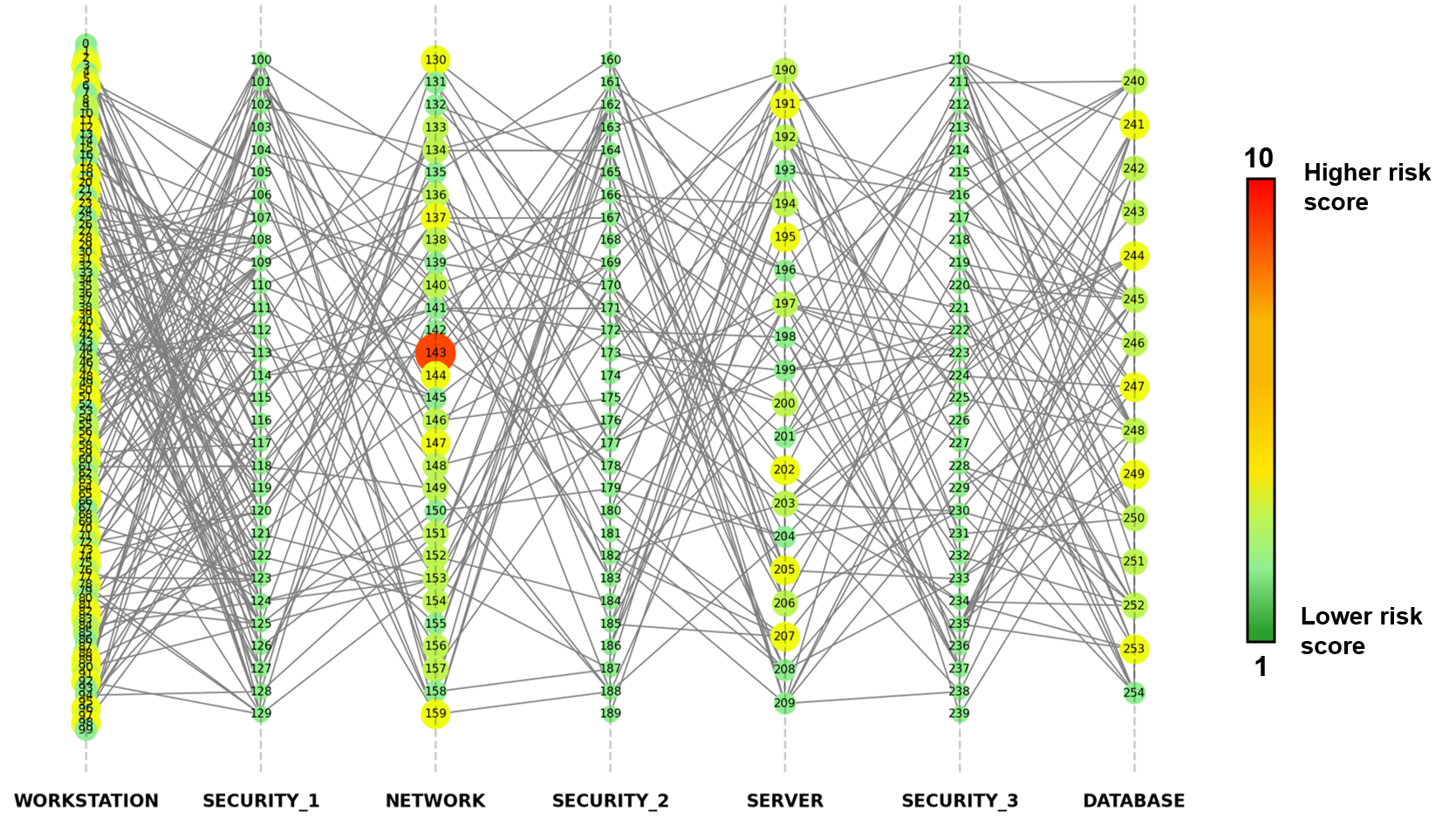}
    \caption{Realistic network topology. Nodes are grouped into seven layers: workstations, network, servers, databases and security layers in between the previous ones. Color and size depend on the initial risk score, while numbers reflect the node identification.}
    \label{fig:big IT}
\end{figure}

Once the weights of the Hamiltonian components were calibrated using the benchmark topology, we applied our QUBO-based risk modeling framework to a more realistic and complex IT infrastructure consisting of 255 individual nodes, organized as follows (Fig.~\ref{fig:big IT}):

\begin{itemize}
    \item 100 Workstations, grouped into 10 isolated subnetworks
    \item 30 Network infrastructure nodes
    \item 20 Servers
    \item 15 Databases
\end{itemize}

The initial risk score for each node was randomly assigned a value between 1 (low risk) and 4 (moderate risk), simulating a relatively secure environment under normal operating conditions. To simulate realistic security architecture and introduce risk mitigation barriers, we inserted a security layer between each pair of adjacent functional layers. These intermediary layers consist of 30 nodes initialized with the minimum risk score. Conceptually, these nodes represent protective systems such as firewalls, intrusion detection/prevention systems (IDS/IPS), network segmentation gateways, or zero-trust access brokers--all intended to inhibit the free propagation of risk between layers.

Regarding connectivity, all nodes of the same type (e.g., network-to-network, server-to-server) were fully connected within their layer, with the exception of workstations belonging to different subnetworks, which remained isolated from each other. Additional inter-layer connections were generated randomly but constrained to occur only between adjacent layers. This ensures a layered structure where, for instance, workstations connect to network devices, but not directly to servers or databases, thereby preserving the intended hierarchy and containment boundaries of the architecture. No cross-layer ``shortcut'' connections were permitted in this initial setup.

To simulate a localized threat scenario, a single node was randomly selected and assigned a high initial risk score ($IS = 8$), representing a critical vulnerability or breach. This allowed us to analyze how risk from a single point of failure propagates through the network under the influence of connectivity and structural layout.

At this stage, no node was flagged as exposed to the internet or missing updates--meaning all update status and exposure flags were set to 0. This constraint allows us to isolate the pure effect of topology and local risk scores on systemic propagation, without confounding variables from known exposure multipliers.

This scaled-up, structured yet randomly instantiated topology offers a valuable middle ground between complete synthetic control and real-world complexity. It enables the study of emergent patterns and model behavior under richer, more varied structural conditions.

\subsection{Problem Formulation}
\label{sec:ProblemFormulation}

The QUBO framework is a flexible mathematical model that enables the encoding of complex decision problems into a binary optimization form \cite{glover2019qubo,kochenberger2014}. In this formulation, each element is associated with a binary variable $x_i \in \{0,1\}$, where the value 1 typically denotes the activation of a condition while 0 indicates its absence. The objective is to minimize a global cost function defined as:

\[
H(\mathbf{x}) = \sum_{i} Q_{ii}x_i + \sum_{i<j} Q_{ij}x_ix_j,
\]

where matrix $Q$ contains the real-valued coefficients encoding both the local characteristics of individual nodes ($Q_{ii}$) and the pairwise interactions between them ($Q_{ij}$). This structure, and in particular the pairwise interactions, allows our model to express intricate dependencies across the network, such as how risk might propagate from one node to another or how clusters of vulnerabilities may amplify overall exposure.

One of the key strengths of the QUBO formulation lies in its compatibility with a wide range of computational approaches, including both classical metaheuristics and quantum solvers \cite{lucas2014}. Moreover, it offers a high degree of modeling flexibility: each characteristic of the nodes in the infrastructure--risk scores, update statuses, connectivity and exposure to the internet--can be mapped to a corresponding mathematical term within the cost function. These terms can then be combined in a coherent manner to reflect the global behavior of the system under study.

To do so, we structure the objective function as a weighted sum of five Hamiltonian components: each component is designed to capture a specific risk-related phenomenon, and their weighted combination yields a total Hamiltonian that reflects the overall vulnerability landscape of the infrastructure.

\begin{equation}
H = \lambda_1 H_1 + \lambda_2 H_2 + \lambda_3 H_3 + \lambda_4 H_4 + \lambda_5 H_5
\end{equation}

Each Hamiltonian $H_k$ encodes a specific risk dynamic, incorporating both domain knowledge and network topology. The weights $\lambda_k$ modulate their relative influence and are carefully tuned based on realistic propagation patterns. Tuning the structure and weights of the Hamiltonians is critical: even minor changes to these components alter the optimization landscape significantly. Our methodology starts from simple, interpretable network topologies to infer which metrics and weight combinations best reflect risk propagation, and generalizes to larger networks where intuition and visual inspection are no longer feasible.

In the following part of this section we break down each term and its contribution to the whole Hamiltonian:

\begin{itemize}

\item $H_1$ --
This Hamiltonian component favors the final risk scores of the nodes staying close to their initial value. It aims to minimize unwanted collective drifts of the risk towards common values and makes it increasingly difficult for nodes with high initial risk to drastically change:

\begin{equation}
H_1 = \sum_i \left( IS_i \cdot (IS_i-FS_i)^2\right)
\end{equation}

where $IS_i$ is the initial score for node $i$ and $FS_i$ is the final risk score of node $i$.

\item $H_2$ --
This term penalizes the final scores of nodes that have many connections, making them inherently riskier:

\begin{equation}
H_2 = -\sum_{i< j} S_{ij} \cdot FS_i \cdot FS_j
\end{equation}

where $S_{ij}$ represents the strength of the connection between node $i$ and node $j$, while $FS_i$ and $FS_j$ are the final scores of nodes $i$ and $j$ respectively.

\item $H_3$ --
This term ensures that the final score is influenced by the score of all the node's neighbors:

\begin{equation}
H_3 = \sum_{i} \left( FS_i-\dfrac{\sum_{j\in N_i}FS_j}{|N(i)|} \right)^2
\end{equation}

where $N_i$ are the nodes connected to node $i$.

\item $H_4$ --
This term penalizes connectivity between nodes marked by critical exposure flags, which might accelerate systemic compromise:

\begin{equation}
H_4 = -\sum_{i<j}  (f_i^\text{no\_update} + f_i^\text{internet} + f_j^\text{no\_update} + f_j^\text{internet}) \cdot (FS_i+FS_j)
\end{equation}

where $f_i^\text{no\_update}$ is 1 if node $i$ is not updated (and thus riskier) and $f_i^\text{internet}$ is 1 if the resource represented by node $i$ is exposed to the internet.

\item $H_5$ --
A final term to isolate and strongly penalize decreases in critical nodes ($IS_i \geq 7$), as in this scenario high-risk nodes tend to remain risky even when surrounded by low-risk nodes:

\begin{equation}
H_5 = -\sum_{i:\ IS_i \geq 7} FS_i
\end{equation}

\end{itemize}

\subsection{Solution Techniques}

To solve our QUBO-based approach, we employed classical, quantum, and hybrid methods.

The overall workflow begins with the construction of the infrastructure graph, where each node is initialized with a risk level and optional flags indicating exposure or patching status. The interactions between nodes, along with structural characteristics such as subnetworks and intermediary security layers, are then encoded into the Hamiltonian. Once all components are combined, the resulting QUBO is submitted to the chosen solver: classical, quantum, or hybrid. The solver returns a binary configuration minimizing the total energy, from which we can extract the final risk profile of the system. This includes identifying which nodes have been affected by propagation, estimating the systemic impact of a critical vulnerability, and pinpointing the most strategic nodes to isolate or reinforce. This integrated approach enables risk to be assessed not merely as a collection of isolated vulnerabilities, but as an emergent property shaped by topology and interdependencies.

All solver workflows--classical, quantum, and hybrid--were executed using D-Wave's cloud-based optimization platform and its corresponding solver suite \cite{Pakin2018}.

\subsubsection*{Classical approach}

On the classical side, we relied on Tabu Search \cite{Glover1989, Glover1990}, a metaheuristic that iteratively explores the solution space while avoiding cycles and local traps by maintaining a memory of recently visited configurations. This memory-based mechanism allows the solver to escape local minima and explore more diverse regions of the landscape. Tabu Search is particularly suited for combinatorial problems like QUBO, where the search space is large and discrete, and it provides a good balance between solution quality and computational efficiency. However, classical approaches tend to scale poorly as the size and density of the QUBO increase. As the energy landscape becomes more rugged and high-dimensional, the probability of getting trapped in local minima grows, and the time required to explore the space increases significantly. This makes purely classical workflows progressively less advantageous for large, densely connected infrastructures.

\subsubsection*{Quantum approach}

On the quantum side, we investigated the use of quantum annealing \cite{Hauke2020, Rajak2022}, a technique designed specifically for solving problems that can be mapped to a QUBO or an equivalent Ising model. Quantum annealing draws inspiration from thermal annealing in metallurgy, where materials are heated and slowly cooled to reach a low-energy configuration. In the quantum analogue, quantum fluctuations, rather than thermal ones, are used to explore the solution space. The process begins with the system initialized in a superposition of all states; the Hamiltonian is then gradually morphed into the target problem Hamiltonian. Through this evolution, the system relaxes toward a low-energy configuration of the problem. Crucially, \emph{quantum tunneling} allows the annealer to transition through energy barriers rather than climbing over them, potentially making it effective in navigating landscapes with sharp local minima. Another advantage of quantum annealing is that the annealing time remains fixed regardless of problem size: the total time to solution depends primarily on the number of shots (repetitions) required, since the process is probabilistic. In theory, this enables better scaling and a higher likelihood of approaching the global minimum compared to classical heuristics. Nonetheless, a significant limitation remains: the QUBO must be embedded onto the restricted connectivity of the Quantum Processing Unit (QPU). In fact, current hardware imposes strict limits on the size and density of solvable problems, making embedding a major bottleneck for highly interconnected cyber-risk QUBOs.

\subsubsection*{Hybrid approach}

To bridge the gap between the flexibility of classical solvers and the exploratory power of quantum annealing, we employed hybrid quantum--classical approaches. In these workflows, most of the optimization is carried out classically, but the solver calls the quantum annealer as a specialized subroutine to tackle the most challenging regions of the energy landscape. In theory, this design leverages the strengths of both paradigms: the classical component provides scalability and robustness, while the quantum component enhances the search process by escaping deep or narrow local minima through tunneling. Because hybrid methods operate on problem fragments rather than the full QUBO, they substantially alleviate the embedding limitations that constrain pure quantum approaches, making it possible to handle larger and more densely connected infrastructures.

Regarding scalability, the hybrid approach behaves differently from the classical and quantum solvers, which both converge after a runtime that is not known a priori. In contrast, the execution time of the hybrid method is explicitly fixed by the user. In principle, this time budget should account for the cost of the classical optimization plus the overhead of sending at least one subproblem to the QPU. To enable a fair comparison across the three approaches, we evaluate the hybrid solver under three timing regimes: the minimum time suggested by the solver, 30 seconds, and 180 seconds. The first case is expected to behave similarly to the classical method, since only a limited number of quantum subroutine calls can be executed. The longer time budgets, however, allow more frequent QPU calls and therefore should benefit more from the quantum solver.

\section{Results}
\label{sec:Results}

\subsection{Classical Simulations}

\begin{figure}[!htbp]
    \centering
    \begin{subfigure}{0.8\linewidth}
        \centering
        \includegraphics[width=\linewidth]{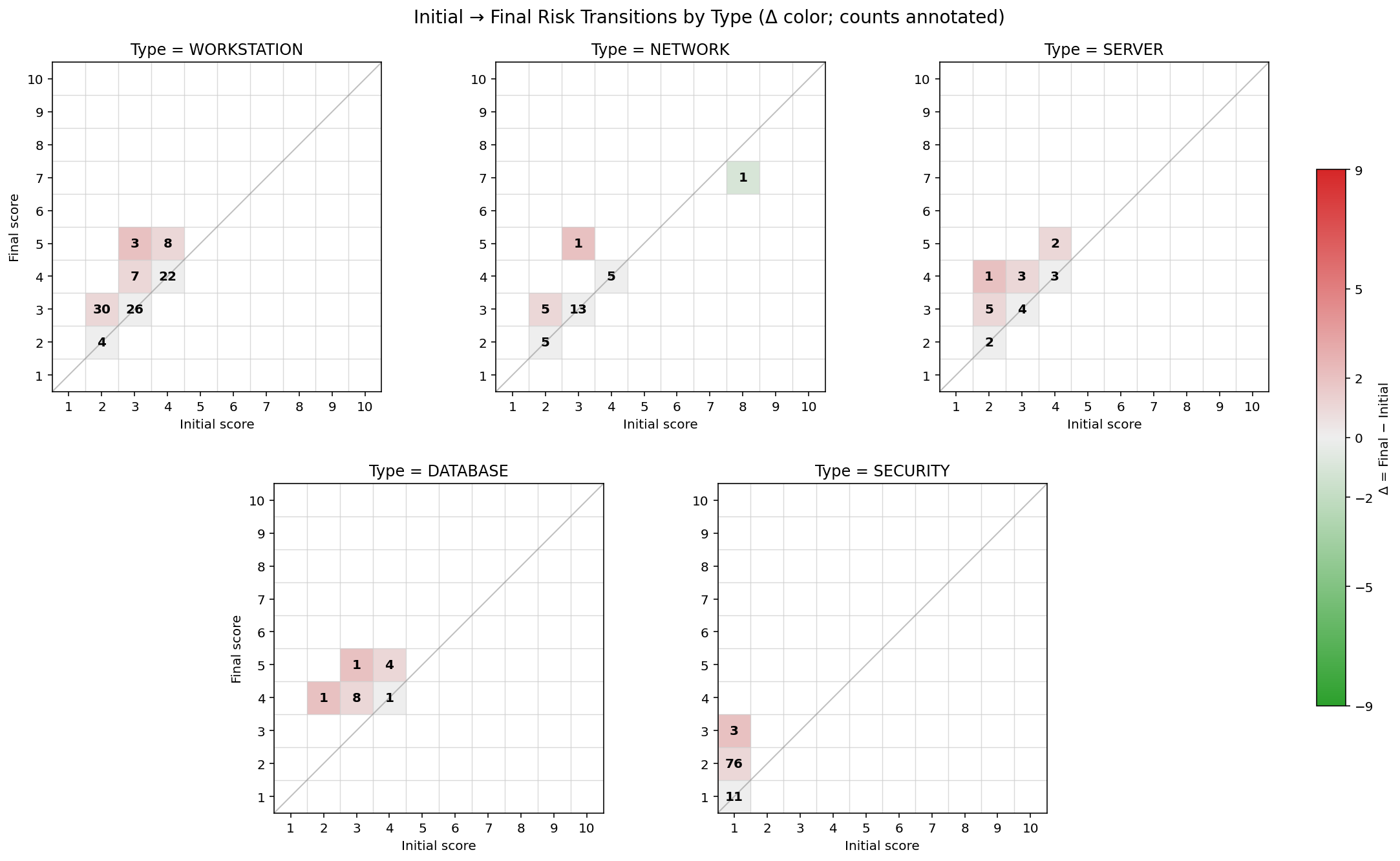}
        \caption{Aggregated results.}
        \label{fig:plain aggr}
    \end{subfigure}

    \vspace{1.5em}

    \begin{subfigure}{0.8\linewidth}
        \centering
        \includegraphics[width=\linewidth]{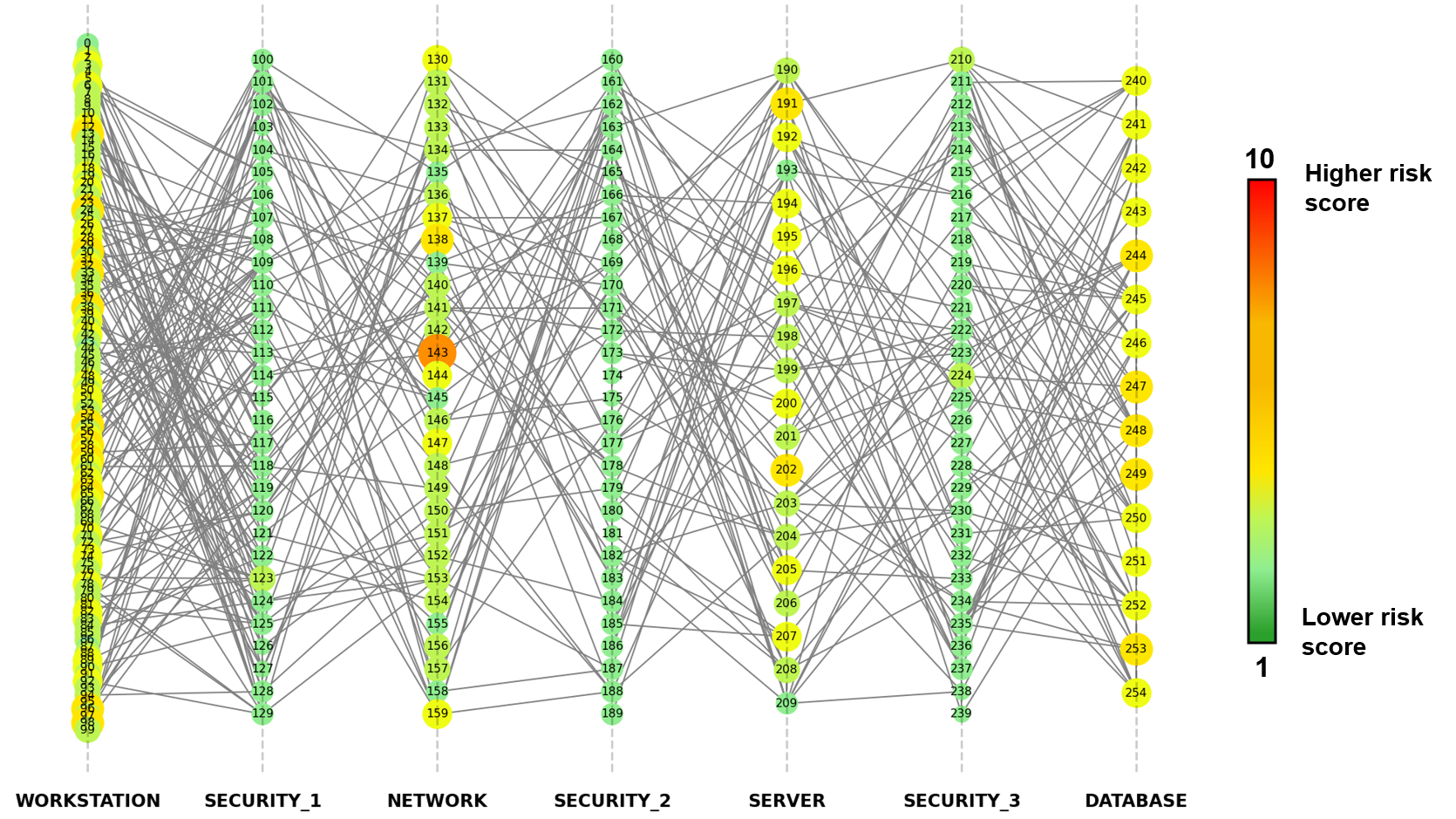}
        \caption{Detailed results.}
        \label{fig:plain detail}
    \end{subfigure}
    \caption{Result of the QUBO minimization on the IT infrastructure in Fig.~\ref{fig:big IT}. Panel (a) shows the aggregated results, depicting for every node type a transition matrix, while (b) shows a detailed snapshot of the network risks after the optimization.}
    \label{fig:qubo_comparison res 1}
\end{figure}

A first way to analyze the outcomes of our QUBO-based cyber risk modeling approach, applied to the infrastructure depicted in Fig.~\ref{fig:big IT}, is by examining the aggregated distributions of node risk scores before and after minimization. The specific initial configuration shown in the figure is just one representative example: the initial scores were assigned randomly, including the selection of the ``risky exception.'' We repeated the experiment for ten different choices of the high-risk node and obtained qualitatively similar results in all cases, confirming that the behavior observed in the displayed configuration is not incidental but characteristic of the model. Figure~\ref{fig:plain aggr} reports comparative transition matrices for all the different infrastructure layers. In each matrix the x-axis represents the initial score and the y-axis the final one; the number populating each entry shows how many nodes for that layer underwent a transition from that initial score to the corresponding final one. The color becomes progressively red the more the transition is towards high risk nodes, and greener in the opposite direction. All the values on the diagonal are nodes that kept their risk score value unchanged.
This visualization provides a macroscopic, aggregated perspective on the system, offering a sense of how the risk distribution shifts across the whole layers.

Performing QUBO using classical simulators, several important trends emerge. First, the average network risk increases slightly, reflecting a mild drift of the global distribution towards higher values. At the same time, the overall dispersion of scores decreases, especially among the same layers. This implies that risk becomes more distributed across the infrastructure: highly vulnerable nodes reduce their exposure, while low-risk nodes absorb part of the burden, leading to a more homogeneous landscape. In practical terms, the model shifts the system away from isolated vulnerabilities towards a smoother distribution of medium-level risk.

The mechanism behind this effect can be traced back to the structure of the Hamiltonians introduced in Section~\ref{sec:Methods}. Highly risky nodes are compensated by the presence of low-risk neighbors, effectively redistributing part of their exposure across the network. Conversely, because of the connectivity-aware penalty term, nodes with many interconnections see their risk score increase slightly, reflecting the intuition that critical nodes--by virtue of their numerous dependencies--are inherently riskier. Together, these dynamics produce a balancing act: risk outliers are dampened, while well-connected nodes inherit additional exposure. The result is a more stable, less heterogeneous network configuration.

This redistribution is visible in the case of the artificially introduced ``risky exception,'' a single node initialized with a critically high score ($IS = 8$). After optimization, its risk is mitigated, but this comes at the cost of increased risk among its neighbors. The presence of security layers between infrastructure tiers prevents this effect from propagating uncontrollably, meaning the global impact of the exception remains contained. This behavior highlights one of the strengths of the QUBO approach: it can capture localized absorption phenomena while simultaneously reflecting global structural constraints.

A complementary mode of analysis is provided by direct visual inspection of the infrastructure topology before and after QUBO minimization depicted in Fig.~\ref{fig:plain detail}. From this perspective, the patterns observed in the matrices are confirmed: the risky exception diminishes in severity, its neighbors grow moderately riskier, and the overall risk profile drifts upwards. However, this effect is difficult to perceive precisely by visual inspection alone as the changes are moderate. While topological visualizations are extremely detailed and intuitively appealing in smaller systems, they quickly become impractical in larger infrastructures, where complexity obscures individual interactions. For this reason, we consider them mainly as an explanatory tool to validate the QUBO model's behavior in interpretable benchmark scenarios, while relying on aggregated statistical representations for larger-scale analyses.

\begin{table}[!htbp]
\centering
\small
\caption{Summary statistics of risk scores before and after QUBO minimization for the 255-node infrastructure.}
\label{tab:summary}
\begin{tabular}{@{}lcccc@{}}
\hline
Scenario & Mean$_\text{init}$ & Mean$_\text{final}$ & Std$_\text{init}$ & Std$_\text{final}$ \\
\hline
Baseline       & 2.24 & 2.94  & 1.17 & 1.04  \\
Stressed       & 2.24 & 3.06 & 1.17 & 1.53 \\
\hline
\end{tabular}
\end{table}

A second experimental setting was designed to stress-test the QUBO model under less standard conditions by artificially increasing the influence of the initially risky node. The underlying IT infrastructure remained the same as before, but in this case we tuned the weights of all edges connecting to and from the critical node, thereby amplifying its ability to spread risk through the network.

\begin{figure}[!htbp]
    \centering
    \begin{subfigure}{0.8\linewidth}
        \centering
        \includegraphics[width=\linewidth]{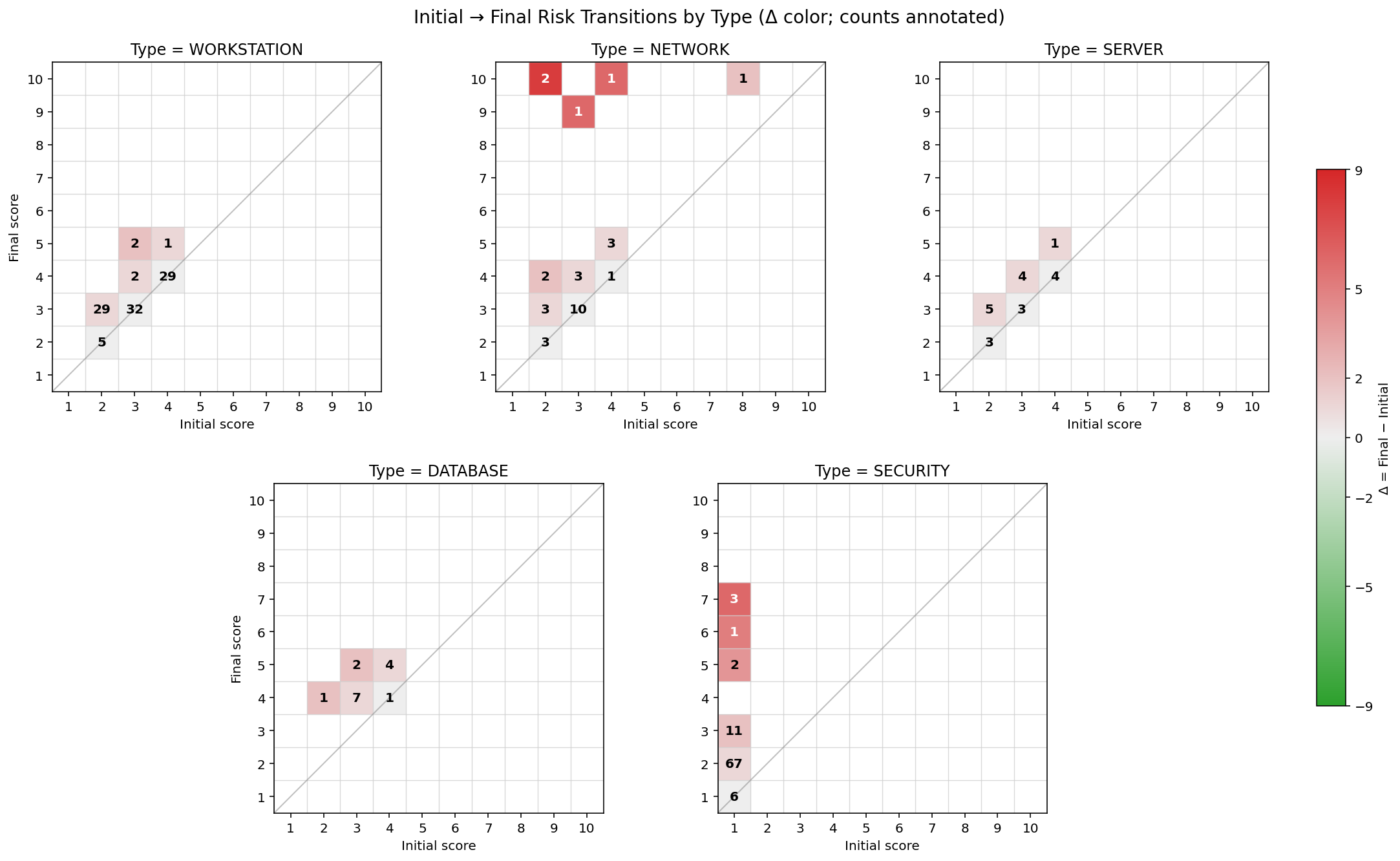}
        \caption{Aggregated results.}
        \label{fig:infl aggr}
    \end{subfigure}

    \vspace{1.5em}

    \begin{subfigure}{0.8\linewidth}
        \centering
        \includegraphics[width=\linewidth]{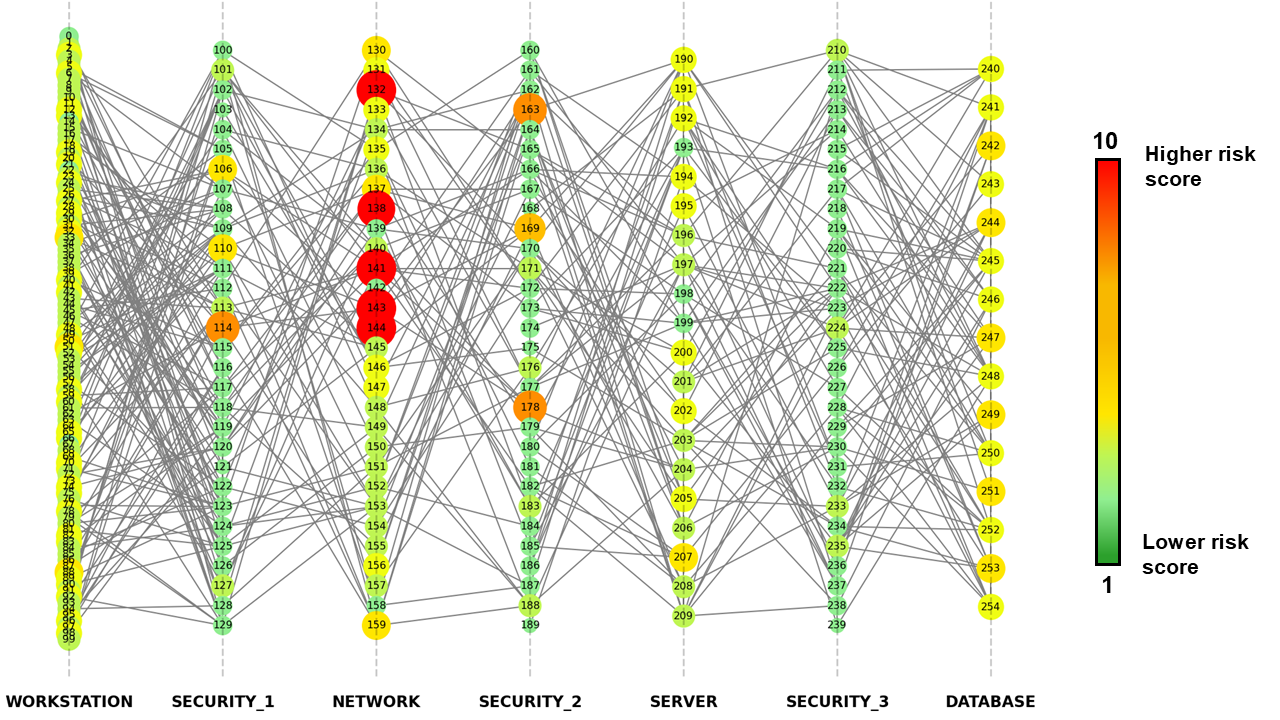}
        \caption{Detailed results.}
        \label{fig:infl detail}
    \end{subfigure}
    \caption{Result of the QUBO minimization after increasing the influence of the high-risk exception (node 143). Panel (a) shows the aggregated results,
depicting for every node type a transition matrix, while (b) shows a detailed snapshot of the network risks after the optimization.}
    \label{fig:qubo_comparison res 2}
\end{figure}

The effect of this modification is immediately visible in both the aggregated and detailed results. Figure~\ref{fig:infl aggr} shows that the distribution of risk scores undergoes a more dramatic shift compared to the baseline scenario: a significant number of nodes accumulate high risk values (with final scores $\geq 7$), while the global average rises more sharply (from 2.24 to 3.06) and the standard deviation increases as well (from 1.17 to 1.53). This indicates that not only did the overall risk drift towards higher values, but the variance across the infrastructure grew, creating new outliers and reinforcing systemic heterogeneity. Unlike the previous case--where risk was redistributed and homogenized--here the network absorbs the shock unevenly, with several nodes emerging as secondary hotspots.

The detailed topology view (Fig.~\ref{fig:infl detail}) confirms this interpretation. The critical node ($\#$143) remains the dominant high-risk element, but its influence clearly permeates beyond immediate neighbors, raising risk scores in multiple areas. This outcome is consistent with the Hamiltonian term penalizing connectivity: nodes linked to a highly influential exception become disproportionately exposed, and because the node was designed to be central, its destabilizing effect cascades through a larger fraction of the infrastructure. Interestingly, some of the nodes that became severely affected are not obvious neighbors of node $\#$143; this illustrates the added value of the QUBO model in uncovering nontrivial propagation paths and latent structural vulnerabilities that would otherwise remain hidden.

\subsection{Quantum and hybrid simulations}

In the following results, we show how classical, quantum, and hybrid solvers differ when applied to our QUBO model. We report how their computational time scales with the size of the infrastructure and evaluate the stability and robustness of their solutions through recursive minimization.

\subsubsection*{Model scalability}

\begin{figure}[!htbp]
    \centering
    \includegraphics[width=0.8\linewidth]{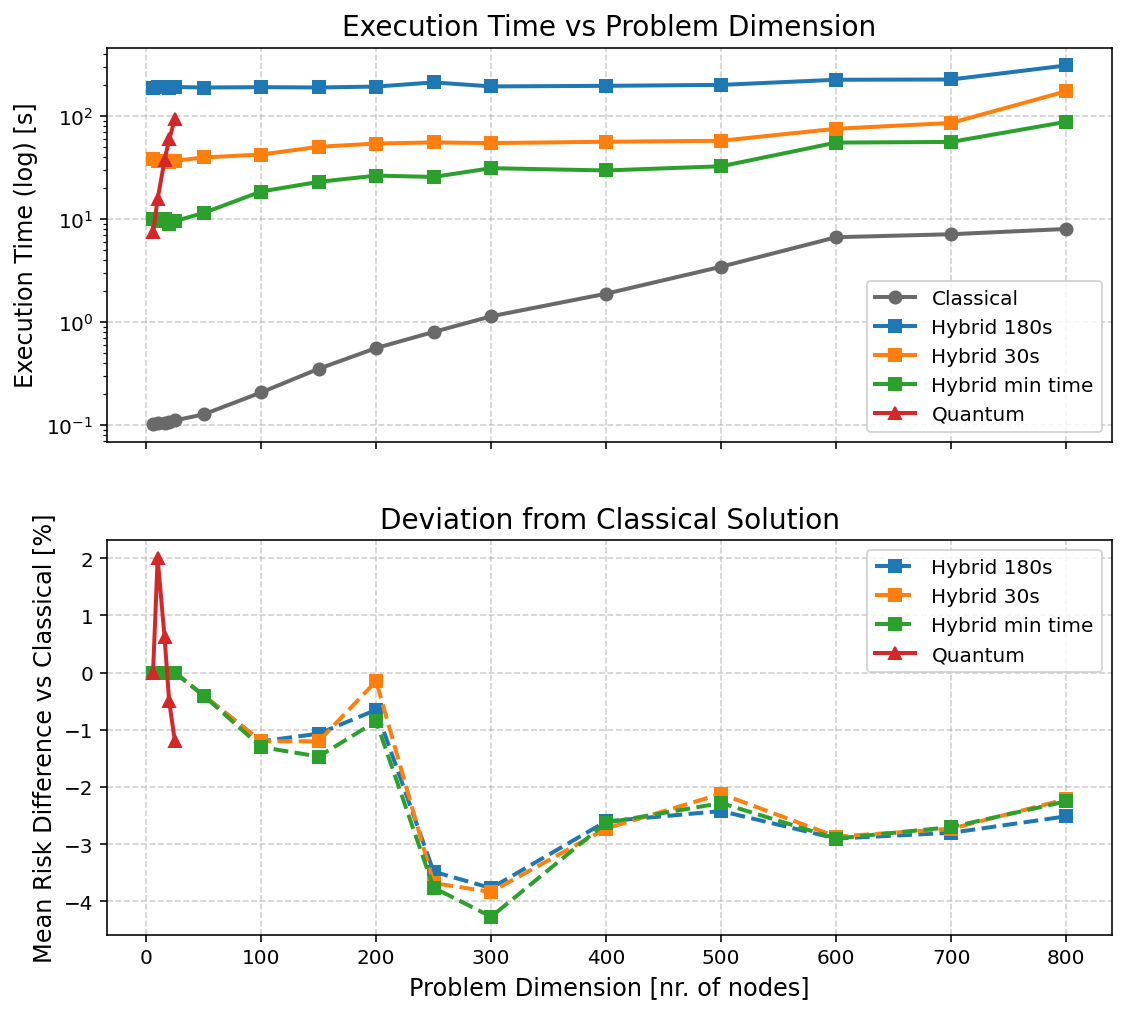}
    \caption{Performance and accuracy comparison across classical, quantum, and hybrid solvers, the latter repeated for three different durations: minimum, 30 seconds, and 180 seconds. The upper plot displays the log-scaled execution time versus problem size, revealing differences in scalability. The lower plot reports the percentage deviation of the quantum and hybrid solutions from the classical reference as the size of the IT infrastructure increases.}
    \label{fig:time_comparison}
\end{figure}

Our first analysis (top part of Fig.~\ref{fig:time_comparison}) compares the computational speed of the three methods--classical, quantum, and hybrid--by measuring the execution time as a function of the problem size. As discussed earlier, classical metaheuristics such as Tabu Search perform efficiently on small and medium-sized instances, but their computational cost increases steeply as the QUBO grows. This trend is clearly visible in the logarithmic plot: the classical curve exhibits an almost exponential growth, appearing nearly linear in log-space. This behavior is consistent with the structure of the cyber-risk formulation. As the number of nodes increases, the density of pairwise interactions grows rapidly, inflating memory requirements and creating a rugged energy landscape with many local minima. Consequently, the cost of exploration rises sharply for large and densely connected infrastructures, making classical solvers progressively less competitive.

The quantum solver displays an even steeper growth pattern and fails to produce solutions beyond approximately 50 nodes. This limitation does not arise from the annealing process itself--which maintains a fixed annealing time and, in principle, stays constant--but from the necessity of \emph{embedding} the problem onto the restricted connectivity of current quantum annealing hardware. Cyber-risk QUBOs are highly interconnected, with edge counts growing rapidly alongside network size. Mapping such dense problems onto architectures such as D-Wave's Pegasus graph \cite{boothby2020} requires complex minor embeddings, a process that is itself NP-hard and often infeasible for larger instances \cite{Lobe2024, Okada2019}. As a result, any potential computational advantage of quantum annealing is overshadowed by the embedding bottleneck, preventing the method from handling realistic infrastructure sizes.

Hybrid solvers offer a more promising trajectory. Because most of the optimization is carried out classically and only selected substructures are delegated to the QPU, hybrid workflows avoid the size limitations faced by pure quantum methods. Their time-to-solution remains largely determined by a user-defined time budget (see Section~\ref{sec:Methods}), aside from a small minimum duration required for at least one quantum subroutine call. In our experiments, hybrid execution times remained stable across problem sizes and aligned with the requested time limits (minimum possible, 30 seconds or 180 seconds). From a scaling perspective, hybrid solvers therefore appear viable for large infrastructures, even if their ultimate performance is still linked to the efficiency of the classical component, and thus might never exceed it. Their competitive behavior suggests that hybrid methods may inherit the classical solver's reliability while benefiting from enhanced exploration through targeted quantum calls.

To compare the solutions produced by each solver in terms of final risk scores, we examined how the final risk distribution compares to the classical baseline. Specifically, we measured the difference between the mean initial risk of each instance and the mean final risk obtained by the classical, quantum, and hybrid workflows across problem sizes. Looking at the bottom part of Fig.~\ref{fig:time_comparison}, remarkably, both the quantum solver (for those small instances where it successfully embeds) and the hybrid solver produce final configurations that deviate from the classical result by only \(-4\%\) to \(+2\%\). This indicates that, for cyber-risk scoring, both quantum and hybrid methods achieve solutions of comparable quality to those of classical heuristics. Moreover, across all tested problem sizes, the hybrid solution consistently produced slightly more conservative final risks--i.e., lower average risk values.

\subsubsection*{Stability of the solution}

In order to assess which method performed best in terms of quality of the risk scoring outcome, we focused on the stability of the obtained solutions. Although QUBO minimization is formulated as a single-step optimization process, we apply it recursively to examine whether repeated minimization drives the system toward stability or instability. In this procedure, the final configuration from one iteration becomes the initial configuration of the next, and the process is repeated multiple times. This technique is not intended as an operational model for cyber risk but rather as a diagnostic tool for evaluating how firmly a solver converges to a stable minimum in the energy landscape.

We applied this recursive minimization for 20 iterations on the 255-node infrastructure shown in Fig.~\ref{fig:big IT} and repeated the experiment for 10 random positions of the initial high-risk node. Due to embedding limitations, the quantum-only solver cannot handle this problem size, so the comparison is restricted to classical and hybrid workflows.

The results reveal a stark contrast. The classical solver consistently exhibits a diverging trend: the mean risk increases across iterations and eventually saturates at the maximum possible value of 10. This behavior indicates that the classical solver often converges to unstable minima, where repeated minimization amplifies risk rather than stabilizing it. In contrast, the hybrid solver frequently identifies stable configurations--states in which repeated minimization produces no further changes. This suggests that hybrid workflows explore the landscape more effectively and are more likely to settle into deeper, more resilient minima. The emergence of these stable basins suggests that hybrid solvers may provide structurally more reliable risk assessments, particularly in large and highly interconnected infrastructures.

\begin{figure}[!htbp]
    \centering
    \includegraphics[width=0.8\linewidth]{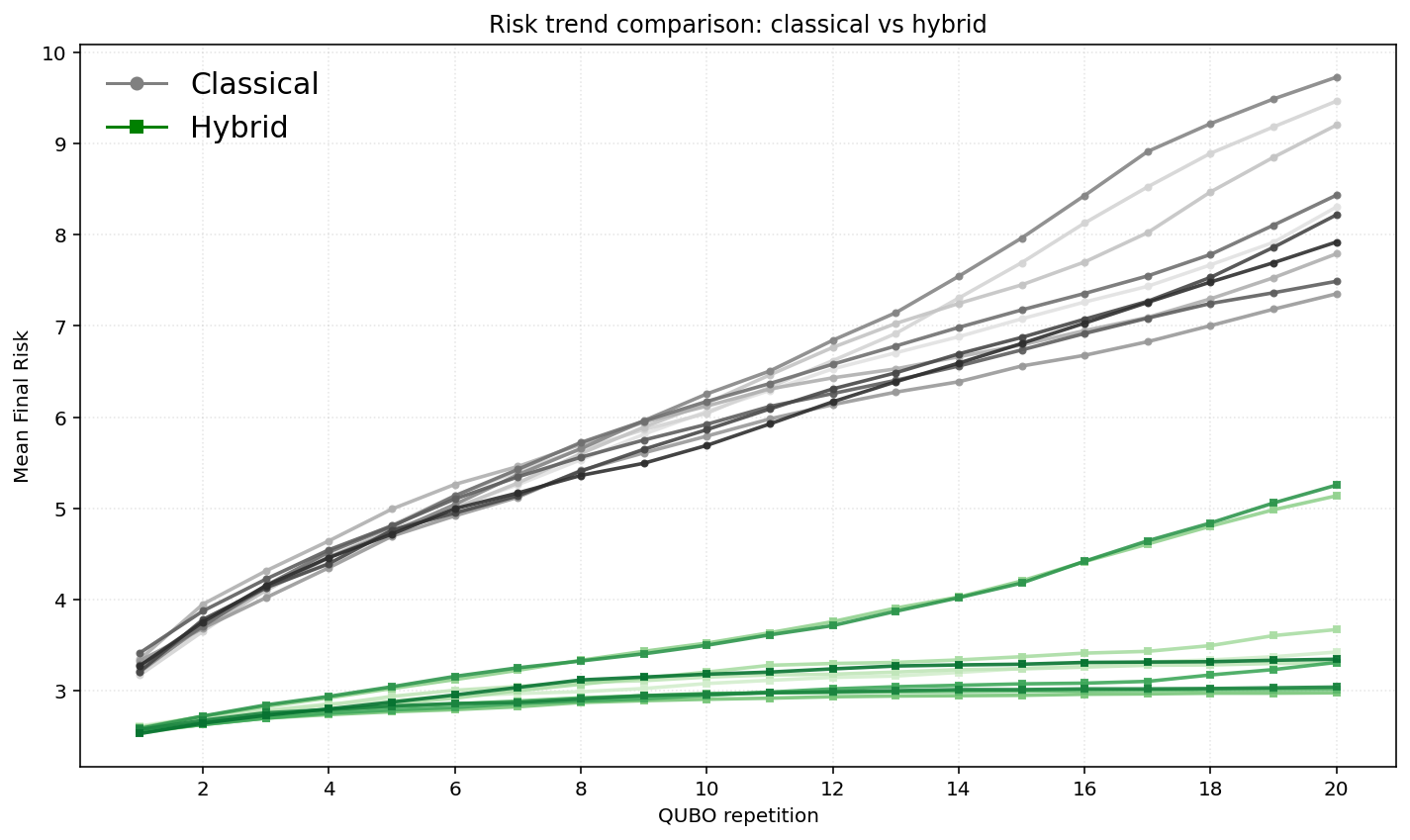}
    \caption{Mean risk trend across iterative application of QUBO on the 255-node IT infrastructure. The different lines represent the same network with the risky exception starting from different random positions. The classical solution (grey) always diverges to maximum risk while the hybrid ones (green) often find a stable solution.}
    \label{fig:recoursive}
\end{figure}

\section{Discussion}
\label{sec:Discussion}

The experimental results highlight how the QUBO formulation captures key aspects of cyber risk dynamics in complex infrastructures. Each Hamiltonian component contributes in a visible and interpretable way: the neighbor-influence term drives local propagation effects, while the connectivity-based penalty increases exposure for highly interconnected nodes, reflecting their critical systemic role. Together, these terms model a realistic spreading of risk inside the infrastructure.

In the first scenario, where a single highly vulnerable node was introduced, the model did not radically transform the system but instead acted as a stabilizing mechanism. The critical node's score was partially absorbed by its neighbors, while surrounding low-risk nodes became slightly riskier in return. This demonstrates the compensatory behavior of the model: instead of leaving isolated vulnerabilities untreated, risk is redistributed across the system in a controlled manner. The presence of intermediary security layers further constrained this effect, confirming their value as effective containment barriers within the infrastructure.

In the second, stressed scenario--where the influential node's connectivity was amplified--the dynamics changed markedly. Risk propagated through non-trivial pathways, giving rise to multiple secondary hotspots and revealing heterogeneous vulnerability patterns. Such patterns are essentially invisible through direct visual inspection of the network, yet the QUBO model exposes them clearly. This illustrates one of the method's main strengths: its ability to uncover latent propagation channels and non-obvious clusters of elevated risk arising from structural interdependencies rather than from local node attributes alone.

Beyond the qualitative analysis of risk distributions, our experiments on solver performance provide important insights into the practical feasibility of QUBO-based cyber risk assessment. Classical metaheuristics such as Tabu Search \cite{Glover1989,Glover1990} remain reliable and efficient at small and medium scales but exhibit sharply increasing runtimes as network density grows. This reflects the expanding number of pairwise interactions and the increasingly rugged optimization landscape. In theory, quantum annealing offers more favorable scaling, as annealing time is fixed and solution time depends primarily on the number of probabilistic shots. However, in practice, the densest cyber-risk QUBOs cannot be embedded efficiently onto the fixed hardware connectivity of current quantum annealers. The embedding step, which is itself NP-hard, becomes the dominant bottleneck and ultimately prevents quantum-only workflows from outperforming classical solvers or even handling larger infrastructures. This finding shows that the theoretical advantages of quantum annealing are, at present, overshadowed by hardware connectivity constraints.

Hybrid quantum--classical workflows emerge as a more promising alternative. Because they offload only selected subproblems to the QPU while retaining a classical outer loop, they avoid the embedding bottleneck and maintain consistent runtime behavior even as problem size increases. More importantly, the recursive-stability experiment indicates that hybrid solvers tend to identify more resilient minima. When the QUBO minimization is applied iteratively, classical solutions frequently diverge with risk escalating across iterations until the entire network saturates at maximum exposure. In contrast, hybrid solutions often remain stable under repeated minimization, suggesting that the solver has converged to a deeper and more robust region of the energy landscape.

From a practical perspective, these findings demonstrate that QUBO is a viable and effective framework for cyber risk scoring. Across all tested solvers--classical, quantum (where embeddable), and hybrid--the model consistently captures the core mechanisms of risk propagation, highlights the systemic impact of highly connected nodes, and reveals non-trivial vulnerability patterns that traditional methods cannot detect. On top of this primary result, several secondary yet important implications emerge. First, the topology of the infrastructure strongly shapes systemic exposure, confirming that central nodes should be prioritized in monitoring and mitigation strategies, consistent with research on influential spreaders in complex networks \cite{kitsak2010}. Second, the experiments validate the protective impact of layered defenses: segmentation and intermediary security controls substantially reduce propagation and dampen systemic amplification. Third, the model uncovers hidden propagation paths and risk clusters, offering insights that would remain invisible under purely qualitative or visual analysis.

Scalability considerations add a complementary operational perspective. Our results show that, under current hardware limitations, only classical and hybrid solvers are suitable for realistic cyber-risk assessments, as pure quantum annealing remains constrained by embedding bottlenecks. Classical solutions remain the fastest option and are well suited for routine assessments where turnaround time is critical. Hybrid solvers, in contrast, offer a compelling trade-off: while slower, they consistently find more stable and conservative minima and therefore become especially valuable when accuracy, robustness, and energy-landscape exploration are more important than raw speed.

\subsection{Limitations}

Several limitations of this work should be acknowledged. First, all experiments were conducted on synthetic network topologies with randomly assigned risk scores. While these topologies reflect realistic architectural patterns, they have not been validated against real-world incident data, actual enterprise network maps, or CVE/CVSS-based parameterizations. Demonstrating the model on production infrastructure data would substantially strengthen its practical credibility.

Second, the recursive minimization procedure used to evaluate solution stability is non-standard. The divergence of classical solutions under iteration may reflect the solver's convergence properties rather than an inherent instability of the solutions themselves, since Tabu Search is designed for single-shot optimization. While this diagnostic procedure provides useful insight into the structure of the energy landscape, its interpretation should be treated with caution.

Third, the scalability experiments reach 800 nodes, which is modest compared to enterprise networks comprising tens of thousands of endpoints. Further work is needed to determine the practical ceiling of the framework and to develop strategies for handling larger-scale infrastructures, potentially through hierarchical decomposition or graph coarsening.

Fourth, all quantum and hybrid experiments were conducted exclusively on D-Wave's annealing platform. The QUBO formulation is in principle compatible with gate-based quantum approaches such as the Quantum Approximate Optimization Algorithm (QAOA), but no comparison with alternative quantum hardware or algorithms was performed.

Finally, the Hamiltonian weight coefficients ($\lambda_1$--$\lambda_5$) were calibrated manually using a benchmark topology. While the resulting parameters generalized well across the tested scenarios, a systematic sensitivity analysis or automated calibration procedure would improve robustness and reproducibility. The specific weight values used in our experiments are not disclosed, as they represent calibration parameters developed through iterative domain-expert tuning. However, the qualitative behavior of the model --- including risk redistribution, propagation patterns, and the relative performance of solvers --- was confirmed to be robust across a range of weight configurations during the benchmark calibration described in Section~\ref{sec:Methods}.

\section{Conclusion}

In this work, we introduced a novel framework for cyber risk assessment based on Quadratic Unconstrained Binary Optimization (QUBO). By encoding risk propagation into the Hamiltonian we created and minimizing it using classical, quantum, and hybrid quantum--classical solvers, we demonstrated how systemic vulnerabilities can be assessed dynamically rather than through static or qualitative heuristics. This marks a significant departure from existing cyber risk scoring practices, which often rely on subjective ratings or isolated snapshots that fail to reflect interdependencies within modern infrastructures.

A central advantage of our approach lies in its flexibility. The network topology, node attributes, and Hamiltonian components can be freely customized, enabling practitioners to construct models that reflect virtually any infrastructure--independent of scale, architecture, or sector. Each Hamiltonian term can be tuned to capture domain-specific effects, while the weighting coefficients allow the behavior of the system to be calibrated precisely. Because of this modularity, the QUBO formulation offers a unifying mathematical language for embedding diverse sources of risk, allowing them to be quantified and optimized consistently across contexts. Its applicability extends naturally beyond the cyber domain, making it a generic framework for modeling complex systems with interdependent risks.

Our results show that the QUBO formulation captures essential systemic behaviors: how networks absorb or amplify local vulnerabilities, how highly connected nodes disproportionately influence global exposure, and how non-trivial propagation pathways arise from structural dependencies rather than local properties. The stability analysis based on recursive minimization further revealed that different solvers converge to qualitatively different minima. Classical solvers often settle into unstable configurations that diverge when re-minimized, whereas hybrid solvers converge more consistently to stable and resilient solutions. This suggests that hybrid workflows may be exploring the optimization landscape more effectively, locating deeper minima that classical heuristics often miss.

Scalability analysis highlights important practical considerations. Classical solvers perform reliably at moderate scales, but their runtime increases sharply as network size and density grow. Quantum annealing, while theoretically promising due to fixed annealing times and superior exploration of rugged landscapes, is currently limited by the need to embed dense QUBOs into hardware with fixed and relatively sparse connectivity. This embedding problem, which is itself NP-hard, dominates the computational cost and prevents quantum-only approaches from outperforming classical ones or handling large cyber-risk models. By contrast, hybrid quantum--classical workflows avoid the embedding bottleneck by delegating only selected substructures to the QPU. As a consequence, hybrid methods scale more gracefully and consistently produce high-quality, stable solutions, making them the most practical quantum-enhanced option available today.

Looking ahead, several promising research directions emerge. Integrating this QUBO framework with real-time monitoring systems could enable continuous and adaptive risk scoring instead of periodic assessments. Broadening the model to include richer infrastructure types--such as cyber-physical systems, distributed cloud architectures, or IoT ecosystems--would further extend its applicability. Incorporating external threat intelligence or dynamic adversarial models may also enrich the Hamiltonian with real-time signals, linking structural vulnerabilities directly to evolving threat activity. Finally, advances in quantum hardware, particularly improvements in qubit count and connectivity, may unlock the full potential of quantum annealing for large-scale cyber risk assessment.

\section*{Declaration of competing interest}
R. Marini is employed by Assicurazioni Generali S.p.A. and R. Arpe provides consulting services to Generali. This research has no connection to their professional roles at Generali, was not funded by Generali, and was conducted independently by the authors. The authors declare no other competing interests.

\section*{Data availability}
The code and data used to produce the results in this paper are available from the corresponding author upon reasonable request.

%% Bibliography
\bibliographystyle{unsrt}
\bibliography{references}

\appendix

\section{Additional Experimental Results}

This appendix presents supplementary analyses that complement the results discussed in the main manuscript.

\subsection{Impact of Exposure and Update Flags}

\begin{figure}[!htbp]
    \centering
    \begin{subfigure}{0.8\linewidth}
        \centering
        \includegraphics[width=\linewidth]{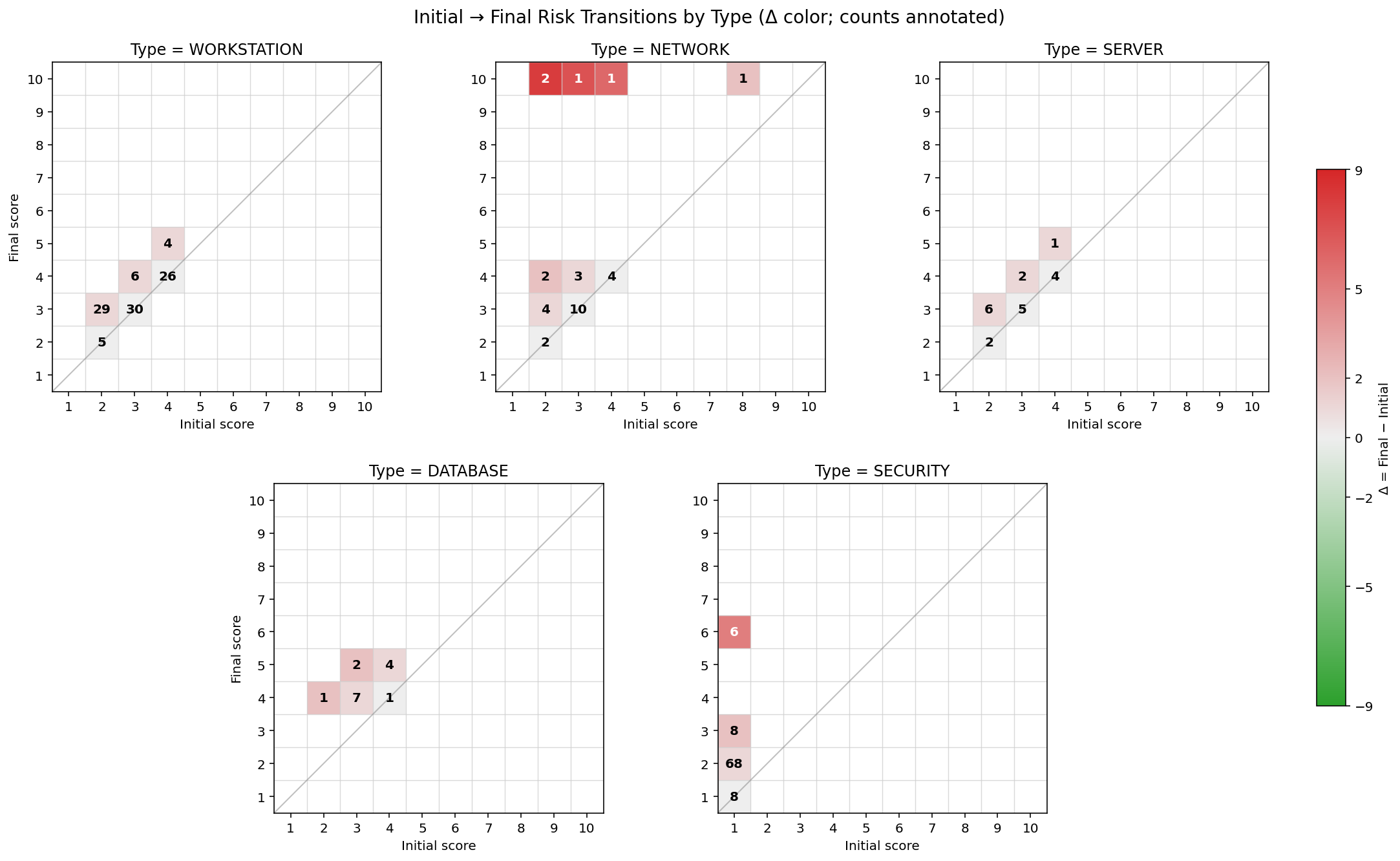}
        \caption{Aggregated results.}
        \label{fig:flag aggr}
    \end{subfigure}

    \vspace{1.5em}

    \begin{subfigure}{0.8\linewidth}
        \centering
        \includegraphics[width=\linewidth]{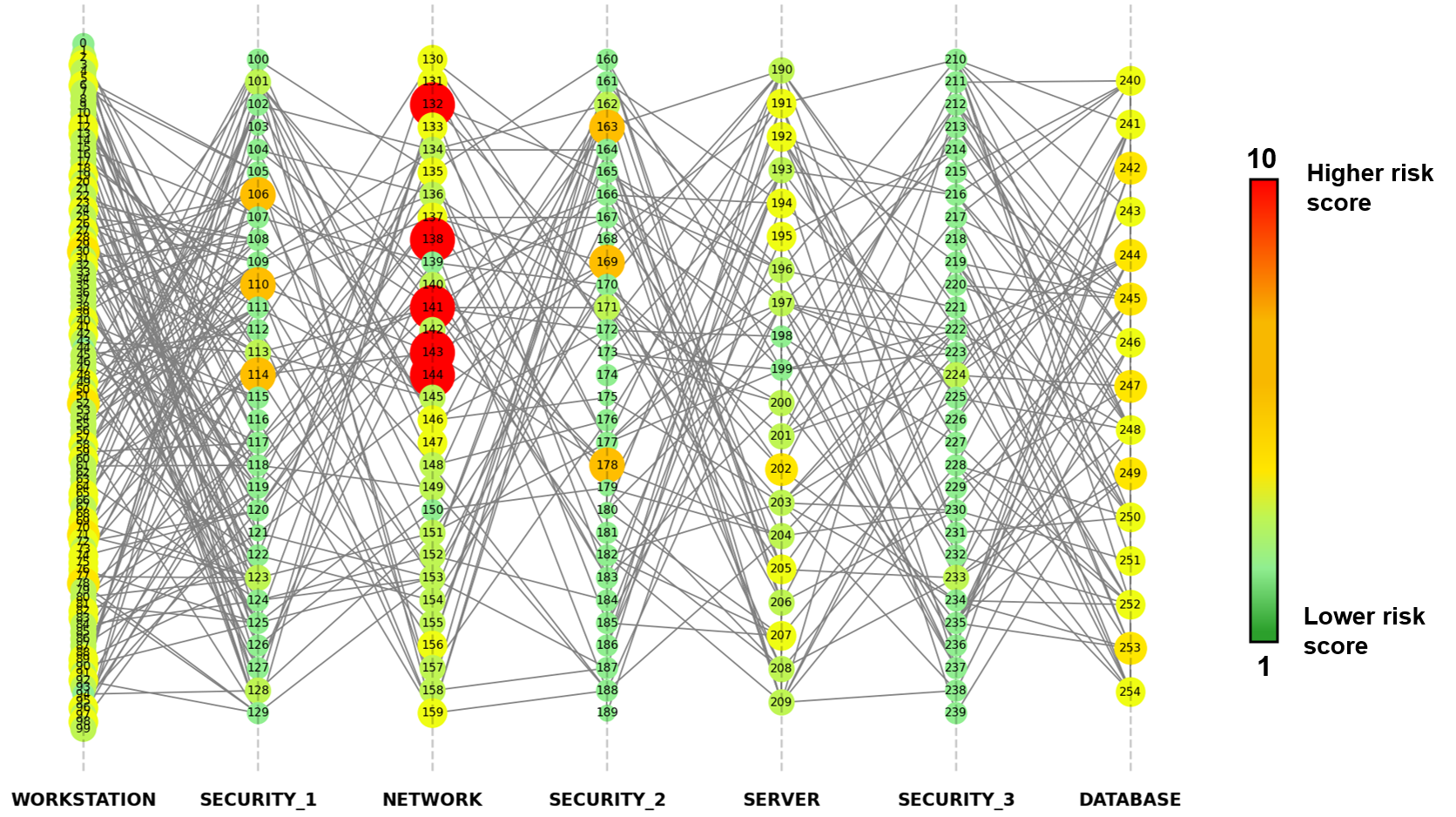}
        \caption{Detailed results.}
        \label{fig:flag detail}
    \end{subfigure}
    \caption{Result of the QUBO minimization after flagging the high-risk exception (node 143) as exposed to the internet and not updated frequently. Panel (a) shows the aggregated results,
depicting for every node type a transition matrix, while (b) shows a detailed snapshot of the network risks after the optimization.}
    \label{fig:flag}
\end{figure}

To further investigate the role of node-level attributes in the QUBO formulation, we examined scenarios in which the initially vulnerable node was also marked as internet-exposed and not regularly updated (Fig.~\ref{fig:flag}). These flags modify the Hamiltonian by increasing both the intrinsic risk contribution of the node and its influence on adjacent components. The resulting configurations show a measurable amplification of systemic risk: exposed or unpatched nodes trigger broader and faster propagation, and their impact often bypasses local containment mechanisms. These experiments confirm that the attribute-driven penalties included in the model behave as intended and meaningfully affect the resulting risk landscape.

The reported results have been obtained with the classical solver, but the same initial setup leads to equivalent results with the hybrid solver, further consolidating the considerations proposed in Section~\ref{sec:Results}.

\subsection{Additional Hybrid Solver Outputs on the 255-Node Infrastructure}

\begin{figure}[!htbp]
    \centering
    \begin{subfigure}{0.8\linewidth}
        \centering
        \includegraphics[width=\linewidth]{normal_results_agg.png}
        \caption{Aggregated results using the classical solver.}
        \label{fig: classicl cmp}
    \end{subfigure}

    \vspace{1.5em}

    \begin{subfigure}{0.8\linewidth}
        \centering
        \includegraphics[width=\linewidth]{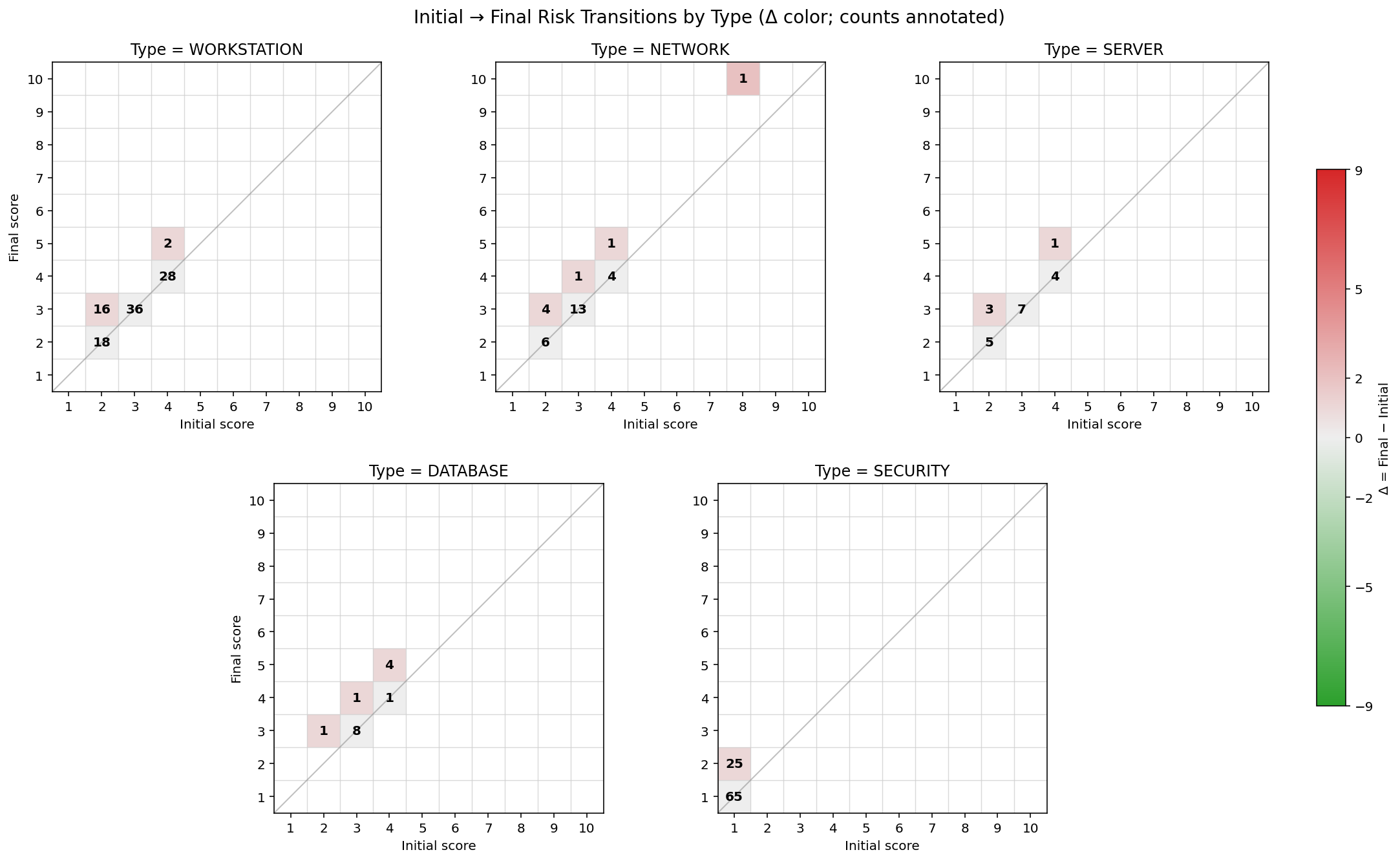}
        \caption{Aggregated results using the hybrid solver.}
        \label{fig:hybrid cmp}
    \end{subfigure}

    \caption{Result of the QUBO minimization on the realistic IT infrastructure depicted in Fig.~\ref{fig:big IT}. Aggregated results showcasing the risk score transitions with the classical solver (top) and the hybrid solver (bottom).}
    \label{fig:comparison}
\end{figure}

For completeness, we also report the results obtained using the hybrid solver on the same 255-node layered infrastructure presented in the main Results section (Fig.~\ref{fig:hybrid cmp}). This allows a direct, side-by-side comparison of how different solution workflows handle the exact same cyber-risk configuration. The figures below show the aggregated risk results produced by the hybrid solver, enabling a transparent assessment of the similarities and divergences with respect to the classical baseline.

\end{document}